\documentclass[11pt]{article}
\usepackage{psfig}
\newcommand{\drline}{\par\unitlength1cm
                     \begin{picture}(14,0.5)
                       \thicklines\put(1.5,0.0){\line(1,0){11}}
                     \end{picture}\par}

\def\thebibliography#1{\section*{} \list
 {[\arabic{enumi}]}{\settowidth\labelwidth{[#1]}\leftmargin\labelwidth
 \advance\leftmargin\labelsep
 \usecounter{enumi}}
 \def\newblock{\hskip .11em plus .33em minus .07em}
 \sloppy\clubpenalty4000\widowpenalty4000
 \sfcode`\.=1000\relax}

\newcounter{saveeqn}

\newfont{\AbsText}{cmr10}
\newfont{\Em}{cmti10}
\newfont{\Bf}{cmbx10}

\newcommand{\boldalpha}{\mbox{\boldmath$\alpha$}}
\newcommand{\boldA}{\mbox{\boldmath$A$}}
\newcommand{\boldr}{\mbox{\boldmath$r$}}

\begin{document}

\setlength{\baselineskip}{14pt}


\begin{center}
  {\Large \bf Dirac--Maxwell Solitons} \\ [18pt]

  {\AbsText \setlength{\baselineskip}{12pt} 
    C. Sean Bohun$^{1)}$ and  F. I. Cooperstock$^{2)}$ \\ [3pt]

    1)
    {\Em Department of Mathematics and Statistics, University of Victoria, \\
      P.O. Box 3045, Victoria, B.C., Canada V8W 3P4 \\
    }
    2)
    {\Em Department of Physics and Astronomy, University of Victoria, \\
      P.O. Box 3055, Victoria, B.C., Canada V8W 3P6 \\ [18pt]
    }
    ABSTRACT \\ [6pt]
    \parbox{22pc}{
      \setlength{\baselineskip}{12pt}
      Detailed analysis of the coupled Dirac-Maxwell equations and the 
      structure of their solutions is presented.
      Numerical solutions of the field equations in the case of spherical 
      symmetry with negligible gravitational self-interaction reveal the 
      existence of families of solitons with electric field dominance
      that are completely determined by the observed charge and mass of the
      underlying particles. A soliton is found which has the charge and mass
      of the electron as well as a charge radius of $10^{-23} m$.  This is
      well within the present experimentally determined upper limit of
      $\simeq 10^{-18} m$.  Properties of these particles as well as 
      possible extension to the work herein are discussed.
    } \\ [18pt]
  }
\end{center}
\begin{center}
1999 PACS numbers: 03.65.Ge, 11.10.Lm, 12.20.Ds
\end{center}


\section{Introduction}
Through the years, a number of authors have attempted to avoid the 
problems inherent in the point-particle model by focussing upon
finite soliton-like structures. Fields interacting non-linearly provide
the binding without invoking any phenomenological elements. Einstein and 
Rosen\cite{einstein} pointed out many years ago that particles 
should be contained  within a field theory and not exist as independent
entities. Rosen\cite{rosen} made considerable progress in implementing
such a program in a gauge-invariant manner by minimally coupling a scalar
field to the Maxwell field.  However, the soliton solutions yielded
negative masses. Later\cite{rosenstock}, neutral quantized particle states
of positive mass were found and a more complicated model invoking up to three scalar fields coupled to the Maxwell field
was shown to be capable of modeling the known massive
leptons\cite{cooperstock}. However, the particles were spinless and the
view then was that a subsequent quantization of the theory would induce spin.

In 1991, one of the present authors\cite{cooperstock2} suggested an
alternative  route to elementary particle modelling, namely as solitons
of Dirac--Maxwell theory. Since Dirac--Maxwell theory had been so successful
in describing electron spin and magnetic moment, predicting the existence
of the positron and refining the energy levels in interacting systems such
as hydrogen, it seemed reasonable that this might successfully extend to
a self-interacting soliton structure to model the elementary particles
themselves. Spin would already exist in such a model via the spinor
structure of the wave function. Shortly thereafter, such solitons were found
and their properties studied\cite{bohun}. A few years later, Lisi\cite{lisi}
independently discovered some of the results in\cite{bohun}. Recently, there
has been a revival of interest in this field and in particular, the issue of
gravitational coupling in the Dirac--Maxwell system has been
considered\cite{finster}. However, there was the misconception
that gravitation was a necessary ingredient for the creation of the soliton.

In this paper, we develop the essential results in\cite{cooperstock2} 
and\cite{bohun} and discuss the role of gravitation in soliton structure.
The experimental inputs are the respective masses of the electron, muon
and tau, their charge and as a constraint, the upper limit to their
size which is $\simeq 10^{-16}$ cm.
The plan of the paper is as follows: in sec.~\ref{gettofield}, we set out the
essential coupled Dirac--Maxwell equations to be solved. The structure of
the Dirac wave function in spherical coordinates is given and particularized
to the case of electric field dominance. The equation is separated in
sec.~\ref{sepofeqn} and we contrast the standard treatment in which a
potential function is imposed such as in the case of hydrogen and the
present case of the soliton where the derivation of the potential is part
of the problem. The formal structure of the potential in terms of the
Green's function is given. It is shown that there do exist spherically 
symmetric potentials for appropriate choices of quantum numbers.

In sec.~\ref{bndrycondts}, the spherically symmetric energy-momentum tensor
is derived. The relationship between the parameters in the Dirac equation
and the physically measured quantities is discussed and the expression for the 
spatial spread of the soliton is given. The various constraints including
singularity avoidance lead to the required boundary conditions for the problem.

In sec.~\ref{theresults}, the results are presented. New variables of convenience
for numerical integration are introduced. The parameters leading to twenty
ground state solitons are listed. It is found that there is a critical
range which leads to solitons within the experimentally observed upper limit
to the size of the electron. Excited states are presented and the mass ratios are
found.

In the final section~\ref{concremarks}, the essential achievements as well as 
the limitations of the results are discussed. It is stressed that the solitons
have been found without the requirement of significant gravitational
interaction and it is conjectured that gravity will be significant for 
Dirac--Maxwell solitons when $e/m \simeq 1$ in units for which $G=c=1$.
In cgs units, this is $2.58 \times 10^{-4}
\ \mbox{esu}\mbox{gm}^{-1}$.  By contrast, the $e/m$ ratio
for the electron is $2.04 \times 10^{21}$ or $5.27 \times 10^{17}
\ \mbox{esu}\mbox{gm}^{-1}$
in cgs units.


\section{Derivation of the Equations}
\label{gettofield}
The field equations are obtained from the Lagrangian of quantum 
electrodynamics\cite{grif}
\begin{equation}
\label{twofive5}
  L = i\hbar c \bar{\psi}\gamma^\mu\partial_\mu\psi - mc^2\bar{\psi}\psi 
  - \frac{1}{16\pi} F^{\mu\nu}F_{\mu\nu} - e\bar{\psi}\gamma^\mu\psi A_\mu
\end{equation}
where $\psi = (\psi_1,\psi_2,\psi_3,\psi_4)^{\mbox{\scriptsize T}}$
is the Dirac spinor, 
$\bar{\psi} = \psi^{\dagger}\gamma^0 = (\psi_1^*,\psi_2^*,-\psi_3^*,-\psi_4^*)$,
$A^\mu = (\varphi,\mbox{\boldmath$ A$})$ is the electromagnetic four-vector
potential and $F^{\mu\nu} = \partial^\mu A^\nu - \partial^\nu A^\mu$ is the 
Maxwell tensor. The $\gamma^\mu$
are $4 \times 4$ Hermitian anticommuting matrices of the unit square
\[
  \gamma^0 = \left(\begin{array}{cc} I & 0\\ 0 & -I \end{array}\right),
  \ \
  \gamma^k = \left(\begin{array}{cc} 0 & \sigma_k \\ -\sigma_k & 0 \end{array}\right),
  \hspace{1cm} k = 1,2,3
\]
where $I$ is the unit $2 \times 2$ matrix and the $\sigma_k$ are the Pauli
matrices
\[
  \sigma_1 = \left(\begin{array}{cc} 0 & 1\\ 1 & 0 \end{array}\right),
  \ \
  \sigma_2 = \left(\begin{array}{cc} 0 & -i\\ i & 0 \end{array}\right),
  \ \
  \sigma_3 = \left(\begin{array}{cc} 1 & 0\\ 0 & -1 \end{array}\right).
\]
Variation with respect to $A_\mu$ and $\bar{\psi}$ respectively,
yield the field equations
\begin{equation}
\label{sno1}
  F^{\mu\nu}_{\ \ ,\nu} = -4\pi\bar{\psi}\gamma^\mu\psi
\end{equation}
\begin{equation}
\label{sno2}
  i\hbar c \gamma^\mu\partial_\mu\psi - mc^2\psi - e\gamma^\mu\psi A_\mu = 0.
\end{equation}
If $\psi$ is chosen to be an energy eigenstate with energy $E$ and one
chooses a static charge distribution with a four-vector potential of the form
\[
  A^\mu = \left( \phi(r,\theta,\varphi),\mbox{\boldmath$ A$}^k(r,\theta,\varphi) \right),
  \hspace{1cm} k = 1,2,3
\]
then the equations~(\ref{sno1})-(\ref{sno2}) are reduced to
\begin{eqnarray}
\label{sno3}
  &~& \hspace{-5cm}
  \left[ -i\hbar c \mbox{\boldmath$\alpha\cdot\nabla$} + \alpha_4 mc^2
  - e\mbox{\boldmath$\alpha\cdot A$}
  + e\phi - E \right] \psi = 0 \\
\label{sno4}
  \nabla^2 \phi &=& -4\pi e \psi^{\dagger}\psi \\
\label{sno5}
  \mbox{\boldmath$\nabla$}\times\left(
  \mbox{\boldmath$\nabla$}\times\mbox{\boldmath$ A$}\right)
  &=& 4\pi e \psi^{\dagger}
  \boldalpha\psi
\end{eqnarray}
where $\alpha^k = \gamma^0 \gamma^k$.

In spherical coordinates, $(x,y,z) = (r\sin\theta\cos\varphi,
r\sin\theta\sin\varphi,r\cos\theta)$, the Dirac wave function has
the structure\cite{bands}
\begin{equation}
\label{thestructure}
  \psi(r,\theta,\varphi)_{\hspace{-1.6cm}\begin{array}{c}
              \scriptstyle ~\\[-2mm] \scriptstyle [j = l+1/2]\end{array}}
  = \left(\begin{array}{c}
  \sqrt{\frac{l-m}{2l+1}} g Y_l^m \\[2mm]
  \sqrt{\frac{l+m+1}{2l+1}} g Y_l^{m+1} \\[2mm]
  - i \sqrt{\frac{l+m}{2l-1}} f Y_{l-1}^m \\[2mm]
  i \sqrt{\frac{l-m-1}{2l-1}} f Y_{l-1}^{m+1}
  \end{array}\right), \ \
  \psi(r,\theta,\varphi)_{\hspace{-1.6cm}\begin{array}{c}
              \scriptstyle ~\\[-2mm] \scriptstyle [j = l-1/2]\end{array}}
  = \left(\begin{array}{c}
  \sqrt{\frac{l+m+1}{2l+1}} g Y_l^m \\[2mm]
  - \sqrt{\frac{l-m}{2l+1}} g Y_l^{m+1} \\[2mm]
  - i \sqrt{\frac{l-m+1}{2l+3}} f Y_{l+1}^m \\[2mm]
  - i \sqrt{\frac{l+m+2}{2l+3}} f Y_{l+1}^{m+1}
  \end{array}\right)
\end{equation}
where $f = f(r)$, $g = g(r)$ and the $\{Y_l^m(\theta,\varphi)\}_{l,m}$ is
the set of orthonormal spherical harmonics defined for
$l  =  0,1,\ldots$, $m  = -l,-l+1,\ldots, l$ and
\begin{equation}
\label{sphereharm}
  Y_l^m(\theta,\varphi) = \sqrt{\frac{2l+1}{4 \pi} \frac{(l-m)!}{(l+m)!}}
  \ P_l^m(\cos \theta) e^{i m\varphi}.
\end{equation}
In addition, $m$ is an integer such that $-j \le m + 1/2 \le j$;
$(m + 1/2)\hbar$ is the $z$-component of the total angular momentum.

Consider the spinor with $j=1/2$, $l=0$ and $m=0$ which implies from the
above representation~(\ref{thestructure})
\begin{eqnarray*}
  4\pi\psi^{\dagger}\psi &=& f(r)^2+g(r)^2 \\
  4\pi\psi^{\dagger}\boldalpha\psi &=&
  2 f(r)g(r)\sin\theta (-\sin\varphi,\cos\varphi,0)^{\mbox{\scriptsize T}}.
\end{eqnarray*}
Resolving equations~(\ref{sno4}) and~(\ref{sno5}) into spherical
coordinates gives
\begin{eqnarray*}
  \nabla^2 \phi &=& -e\left(f(r)^2+g(r)^2\right) \\
  \mbox{\boldmath$\nabla$}\times\left(
  \mbox{\boldmath$\nabla$}\times\mbox{\boldmath$ A$}\right)
  \Big|_{\hat{r}} &=& 0 \\
  \mbox{\boldmath$\nabla$}\times\left(
  \mbox{\boldmath$\nabla$}\times\mbox{\boldmath$ A$}\right)
  \Big|_{\hat{\theta}} &=& 0 \\
  \mbox{\boldmath$\nabla$}\times\left(
  \mbox{\boldmath$\nabla$}\times\mbox{\boldmath$ A$}\right)
  \Big|_{\hat{\varphi}} &=& 2 e f(r)g(r)\sin\theta.
\end{eqnarray*}
Therefore, a four-vector potential of the form
\[
  A^\mu = \left(
  \phi(r),
  -A_{\varphi}(r,\theta)\sin\varphi,
  A_{\varphi}(r,\theta)\cos\varphi, 0\right)
\]
should be chosen where the components satisfy
\begin{equation}
  \label{sno6}
  \frac{d^2\phi}{dr^2} + \frac{2}{r}\frac{d\phi}{dr} =
  -e\left( f(r)^2+g(r)^2 \right)
\end{equation}
\begin{equation}
  \label{sno7}
  \frac{\partial^2 A_{\varphi}}{\partial r^2} +
  \frac{2}{r}\frac{\partial A_{\varphi}}{\partial r} +
  \frac{\cot\theta}{r^2}\frac{\partial A_{\varphi}}{\partial\theta} +
  \frac{1}{r^2} \frac{\partial^2 A_{\varphi}}{\partial \theta^2} -
  \frac{A_{\varphi}}{r^2 \sin^2\theta} = -2 e f(r)g(r) \sin\theta.
\end{equation}
Since the right hand side of equation~(\ref{sno7}) is nonzero, the theory
can only be exact if $A_{\varphi}$ is nonzero. However at this point we
will impose the assumption of electric field dominance and hence the
dominance of $\phi$ over $\boldA$ or $f(r)$ dominance over $g(r)$.

For the validity of the approximation $\boldA = \mbox{\bf 0}$,
one radial component of the spinor must dominate over the other so that
\[
  f g \ll f^2 + g^2.
\]
It will be demonstrated that such objects do exist within the non-linear
field. With this approximation the equations to solve reduce to a Dirac
equation coupled to a Poisson equation:
\begin{equation}
\label{dirac1}
  \left[ -i\hbar\mbox{\boldmath$\alpha\cdot\nabla$} + \alpha_4 mc^2
  + V(r)\right] \psi = E \psi
\end{equation}
\begin{equation}
  \label{newpotential}
  \nabla^2 V = -4 \pi e^2 \psi^\dagger \psi.
\end{equation}

With these facts in mind, we now turn to the separation of the stationary
Dirac equation~(\ref{dirac1}) with respect to a general central potential
and the derivation of the form of $\psi^\dagger \psi$ for a general set
of quantum numbers.


\section{Separation of the Equation}
\label{sepofeqn}
The separation procedure follows that given in Bethe and Salpeter\cite{bands}.
First one introduces quantum numbers $l$ and $j$; $l$ is the orbital angular
momentum quantum number as well as being an integer $\ge 0$; $j$ is the total
angular momentum quantum number and can assume just the two values $l + 1/2$
and $l - 1/2$, (but only $+1/2$ for $l = 0$).  The forms assumed by the four
components of $\psi$ are given explicitly in~(\ref{thestructure}).

The explicit form of the Dirac equation~(\ref{dirac1}) for the four
components of the wave function is:
\begin{eqnarray}
\label{starray}
  \frac{\partial \psi_3}{\partial z} + \frac{\partial \psi_4}{\partial x}
  - i \frac{\partial \psi_4}{\partial y}
  - \frac{i}{\hbar c} \left[ E - V(r) - mc^2 \right] \psi_1 &=& 0 \\
  \frac{\partial \psi_4}{\partial z} - \frac{\partial \psi_3}{\partial x}
  - i \frac{\partial \psi_3}{\partial y}
  + \frac{i}{\hbar c} \left[ E - V(r) - mc^2 \right] \psi_2 &=& 0 \\
  \frac{\partial \psi_1}{\partial z} + \frac{\partial \psi_2}{\partial x}
  - i \frac{\partial \psi_2}{\partial y}
  - \frac{i}{\hbar c} \left[ E - V(r) + mc^2 \right] \psi_3 &=& 0 \\
  \label{earray}
  \frac{\partial \psi_2}{\partial z} - \frac{\partial \psi_1}{\partial x}
  - i \frac{\partial \psi_1}{\partial y}
  + \frac{i}{\hbar c} \left[ E - V(r) + mc^2 \right] \psi_4 &=& 0.
\end{eqnarray}
Therefore, by inserting the assumed wave functions~(\ref{thestructure})
into~(\ref{starray})-(\ref{earray}) and using identities 
similar to~(\ref{bad1}) we find that the following two coupled equations
between $f$ and $g$ hold:
\begin{eqnarray}
\label{fin1}
  \frac{1}{\hbar c} \left[ E - V(r) + mc^2 \right]f(r)
  - \left[\frac{dg}{dr} + \frac{1+\kappa}{r} g(r) \right] &=& 0 \\
  \label{fin2}
  \frac{1}{\hbar c} \left[ E - V(r) - mc^2 \right]g(r)
  + \left[\frac{df}{dr} + \frac{1-\kappa}{r} f(r) \right] &=& 0
\end{eqnarray}
where the new quantum number $\kappa$ is defined as
\begin{equation}
\label{defofkappa}
  \kappa = \left\{ \begin{array}{l@{\quad \mbox{for} \quad} ll}
            -l-1 & j = l+1/2 & (l=0,1,\ldots)\\
            l    & j = l-1/2 & (l=1,2,\ldots).
              \end{array} \right.
\end{equation}
These equations are valid for all spherically
symmetric potentials $V(\boldr)=V(r)$ and together they
replace expression~(\ref{dirac1}).

At this point, the standard procedure is to specify an external
spherically symmetric potential, an example of which is the
electrostatic potential energy of the proton-electron interaction.
That is, simply
\[
  V(r) = - \frac{Z e^2}{r},
\]
which is the fundamental solution of Laplace's equation\cite{evans}
\begin{equation}
\label{nostructpart}
  \nabla^2 V = 4\pi Z e^2 \delta^3(\boldr)
\end{equation}
where $\delta^3(\boldr)$ is a three dimensional Dirac delta function
centered at the origin. This is consistent with the far range\footnote{By
far range, we mean those distances much larger than the Bohr radius
$r \gg \hbar^2/me^2$.} behaviour that we expect to find for the self-field
of the fermion since, when we compare~(\ref{nostructpart}) 
with~(\ref{newpotential}) we see that the fermion is treated as an
object without structure through the equality,
\[
  \psi^\dagger \psi = -\delta^3(\boldr).
\]

There is one additional problem that must be explored, namely how to
couple relation (\ref{sno4}) to~(\ref{fin1})-(\ref{fin2}).  This will be
achieved in three parts.  First, we find the Green's function for the
equation~(\ref{sno4}).  Second, we find an analytic form for the
probability density $\psi^{\dagger}\psi$ using~(\ref{thestructure}).
Once this equation is known, we can proceed to the third step which is
to find $V(r)$ by forming the convolution of the Green's function of
step one, with the probability density of step two.

The potential $V(r)$ satisfies the Poisson equation~(\ref{newpotential})
and by assuming that the solution is sufficiently regular, this can
be converted to an integral equation\cite{brpm}
\begin{equation}
\label{greenpart1}
  V(\boldr) = -4 \pi e^2 \int
  G(\boldr,\boldr')
  \psi^\dagger(\boldr') \psi(\boldr') \, d \boldr'
\end{equation}
where $G(\boldr,\boldr')$ is the Green's function
of the Laplacian operator in three dimensions
\begin{equation}
\label{greenpart2}
  G(\boldr,\boldr')
  = -\frac{1}{4\pi} \frac{1}{\left|\boldr-\boldr'\right|}
  = -\sum_{l=0}^{\infty} \frac{1}{2l+1} \frac{r_{<}^l}{r_{>}^{l+1}} 
  \sum_{m=-l}^{l} Y_l^m(\theta,\varphi) Y_l^{m*}(\theta',\varphi').
\end{equation}

With the Green's function determined, we can turn our attention to the
probability density.  This is accomplished by using a pair of
identities for the associated Legendre functions\footnote{Both
Eqs.~(\ref{stident1})-(\ref{stident2})
follow directly from Eqs.~(8.5.1) and~(8.5.3) of
Abramowitz \& Stegun\cite{abram}.}
\begin{eqnarray}
\label{stident1}
  (1-\mu^2)\left(P_l^{m+1}\right)^2
  &=& \left[(l-m)\mu P_l^m - (l+m)P_{l-1}^m\right]^2, \\
  \label{stident2}
  (1-\mu^2)\left(P_{l-1}^{m+1}\right)^2
  &=& \left[(l+m)\mu P_{l-1}^m - (l-m)P_l^m\right]^2
\end{eqnarray}
together with the definition of the spherical harmonics~(\ref{sphereharm}).
The resulting expression for the charge density of the Dirac particle
is given by
\begin{equation}
\label{psirhoplus}
  \psi^{\dagger}\psi = \frac{f^2+g^2}{2l+1} 
  \left[ (l-m) \left| Y_l^{m+1} \right|^2
  + (l+m+1) \left| Y_l^m \right|^2 \right]
\end{equation}
when $j = l + 1/2$ and
\begin{equation}
\label{psirhominus}
  \psi^{\dagger}\psi = \frac{f^2+g^2}{2l+1} 
  \left[ (l+m+1) \left| Y_l^{m+1} \right|^2
  + (l-m) \left| Y_l^m \right|^2 \right]
\end{equation}
when $j = l - 1/2$.

Therefore by using~(\ref{greenpart1}), (\ref{greenpart2})
and~(\ref{psirhoplus}), one obtains the expression
\begin{eqnarray*}
  V(\boldr) &=&
  -4\pi e^2 \int
  G(\boldr,\boldr')\,\psi^{\dagger}(\boldr')\psi(\boldr')\,d\boldr' \\
  &=& \frac{4\pi e^2}{2l' + 1} \int
  \sum_{l=0}^{\infty} \frac{r_{<}^l}{r_{>}^{l+1}}
  \sum_{m=-l}^l \frac{1}{2l+1} Y_l^m(\theta,\varphi) Y_l^m(\theta',\varphi')
  \left[ f(r')^2 + g(r')^2 \right] \\
  &~& \hspace{-1cm} \times \left[ (l'-m') \left| Y_{l'}^{m'+1}(\theta',\varphi')
  \right|^2 + (l'+m'+1) \left| Y_{l'}^{m'}(\theta',\varphi') \right|^2 \right]
  \,r'^2 dr' \, d(\cos\theta')\,d\varphi'
\end{eqnarray*}
for the case $j = l + 1/2$.  
Similarly, with the use of~(\ref{psirhominus}), it can be shown that the
potential $V(\boldr)$ takes the form
\begin{eqnarray*}
  V(\boldr) &=&
  \frac{4\pi e^2}{2l' + 1} \int
  \sum_{l=0}^{\infty} \frac{r_{<}^l}{r_{>}^{l+1}}
  \sum_{m=-l}^l \frac{1}{2l+1} Y_l^m(\theta,\varphi) Y_l^m(\theta',\varphi')
  \left[ f(r')^2 + g(r')^2 \right] \\
  &~& \hspace{-1cm} \times \left[ (l'+m'+1) \left| Y_{l'}^{m'+1}(\theta',\varphi')
  \right|^2 + (l'-m') \left| Y_{l'}^{m'}(\theta',\varphi') \right|^2 \right]
  \,r'^2 dr' \, d(\cos\theta')\,d\varphi'
\end{eqnarray*}
for the case $j = l - 1/2$. It is to be noted that the primed indices $(l', m')$
correspond to the angular momentum of the particle, while the unprimed indices
run over the complete set of permissible angular momentum quantum numbers.
By performing the angular integration of the above formulae, one can
immediately conclude that both of the above integrals vanish except 
when $m = 0$ and $l = 0, 2 ,\ldots, 2l'$.  This implies that
\begin{eqnarray}
  V(\boldr) &=& 
  4\pi e^2 \frac{2(l'-j')}{2l'+1}
  \sum_{n=0}^{l'} \, \frac{Y_{2n}^0(\theta,\varphi)}{4n+1}
  \int_{r' = 0}^{\infty}
  \frac{r_{<}^{2n}}{r_{>}^{2n+1}}\,
  \left[ f(r')^2 + g(r')^2 \right]\,r'^2 dr' \nonumber \\
  \label{finalform}
  &~& \hspace{-1cm} \times
  \left[ (\kappa'+m'+1) \langle l',m'+1|Y_{2n}^0|l',m'+1\rangle
  + (\kappa'-m') \langle l',m'|Y_{2n}^0|l',m'\rangle \right]
\end{eqnarray}
where the cases $j = l \pm 1/2$ have been combined by the application of the
definition of $\kappa'$. Expression~(\ref{finalform})
replaces the equation~(\ref{newpotential}). When written in this form, it is
clearly seen that the potential $V(\boldr)$ is not in general
spherically symmetric.
Table~\ref{tab3} lists the potential~(\ref{finalform}) for $l' = 0, 1$ and
illustrates the fact that there exists spherically symmetric states with
$l' \ne 0$.

A localized solution of this model must satisfy the field
equations~(\ref{fin1}) and~(\ref{fin2}) for $f$ and $g$ and a given
energy $E$ where the potential is given by the expression~(\ref{finalform}).
Moreover, it is required that the total probability
\[
  \langle\psi|\psi\rangle = \sum_{i=1}^4 \langle\psi_i|\psi_i\rangle
  = \int_0^{\infty} \left( f^2 + g^2 \right)\,r^2 dr < \infty.
\]

Since the equations which describe the spatial evolution of the
wave function~(\ref{fin1})-(\ref{fin2}) were derived under the assumption
that the potential, $V(r)$, is spherically symmetric, they are not valid
for an extended Dirac particle in an arbitrary state of angular momentum.
We have shown that there do exist certain choices of $l$ and $m$ where 
the probability density is spherically symmetric and it is these cases in
which our primary interest lies.

We can conclude that with the spinor representation given by~(\ref{thestructure}),
there are essentially three differential equations to be solved simultaneously.
Equations~(\ref{fin1})-(\ref{fin2}) specify the spatial evolution of the
wave function and equation~(\ref{finalform}) reflects the spatial extent of
the self-field of the particle. A strategy for solving these intrinsically
non-linear equations, as well as a few of their interesting properties,
will be explored in the following sections.


\section{Boundary Conditions}
\label{bndrycondts}

From the previous section we have found that the equations to be satisfied
for a self-interacting fermion are equations~(\ref{fin1})-(\ref{fin2}) and
\begin{equation}
  \label{twofive3}
  \nabla^2 V = -4 \pi e^2 \frac{2(l-j)}{2l+1}
  \left( f^2 + g^2 \right) 
  \left[ (\kappa+m+1) \left| Y_l^{m+1}(\theta,\varphi) \right|^2
  + (\kappa-m) \left| Y_l^m(\theta,\varphi) \right|^2 \right]
\end{equation}
where we have combined the $j = l \pm 1/2$ cases by using the definition
of $\kappa$.  Since we have assumed that the potential $V$ in
equations~(\ref{fin1})-(\ref{fin2}) is spherically symmetric,
this necessarily restricts the choice of $l$ and $m$.  Assume from this
point on that $l$ and $m$ are chosen to satisfy this criterion.  Consequently, 
equation~(\ref{twofive3}) becomes
\begin{equation}
\label{twofive4}
  \nabla^2 V = - e^2 \left( f^2 + g^2 \right).
\end{equation}

Since the soliton asires as a coupling between Dirac and Maxwell fields,
the energy $E$ that appears in the Dirac equation is not the total energy
of the particle.  The total energy can be obtained by calculating the
$T_0^0$ component of the energy-momentum tensor.  For our field, 
the Lagrangian is given by equation~(\ref{twofive5})
where $A^\mu$ is the vector potential of the electromagnetic field.
One generates the symmetric energy-momentum tensor directly from the
Lagrangian in the form\cite{landl}
\begin{equation}
\label{twofive6}
  T^{\mu\nu} = \frac{\partial L}{\partial g_{\mu\nu}} - \frac{g^{\mu\nu}}{2} L.
\end{equation}
Applying~(\ref{twofive6}) to~(\ref{twofive5}) yields
\begin{eqnarray*}
  T^{\mu\nu} &=&
  \left[ \frac{i\hbar c}{2} \left( \bar{\psi} \gamma^\mu \partial^{\nu} \psi +
  \bar{\psi} \gamma^\nu \partial_\mu \psi \right)
  - \frac{1}{4\pi}F^{\alpha\mu}F^{\beta\nu}g_{\alpha\beta}
  - \frac{e}{2} \left( \bar{\psi}\gamma^\mu\psi A^\nu
  + \bar{\psi}\gamma^\nu\psi A^\mu \right) \right] \\
  &~& - \frac{g^{\mu\nu}}{2}\left[ i\hbar c \bar{\psi} \gamma^{\alpha}
  g_{\alpha\beta} \partial^{\beta} \psi
  - mc^2 \bar{\psi}\psi - \frac{1}{8\pi} F^{\alpha\beta}F_{\alpha\beta}
  - e \bar{\psi}\gamma^{\alpha}\psi A^{\beta} g_{\alpha\beta} \right].
\end{eqnarray*}
Further simplification gives
\[
  T_0^0 = E \psi^{\dagger} \psi
  + \frac{1}{8\pi} \left( \frac{d\phi}{dr} \right)^2.
\]
This yields an expression for the total energy,
$E_{\mbox{\scriptsize tot}}$, of
\begin{eqnarray}
\label{emident}
  E_{\mbox{\scriptsize tot}} &=& \int T_0^0 \, d V_{ol} \nonumber \\
  &=& E \int_0^\infty \left(f^2+g^2\right) r^2 \, d r
  + \frac{1}{2} \int \left( \frac{d\phi}{d r} \right)^2 \, r^2 \, d r
\end{eqnarray}
where $d V_{ol}$ is an infinitesimal volume element.
This total energy should be associated with the observed mass of the
particle as $E_{\mbox{\scriptsize tot}} = m c^2$.
There is still sufficient freedom remaining to set $\lim_{r\to\infty}V(r)=0$
because the spinor is invariant under the transformation 
\[
  V \to V + \beta; \ \ \ \ \ E \to E + \beta
\]
for any real-valued $\beta$.

The mass $m$ and the charge $e$ that appear in the Dirac equation are not
necessarily the experimentally measured quantities just as the charge that
appears at a vertex of a Feynman graph is not the experimentally measured 
charge of the particle.  Because of this, we will replace the $m$
in~(\ref{fin1})-(\ref{fin2}) by the symbol $\mu$. In addition,
the $e$ in~(\ref{twofive4}) will be replaced by an $\epsilon$.
The symbols $m$ and $e$ will be reserved for the physically 
observed quantities. With these substitutions, we convert to a set of
variables whereby equations~(\ref{fin1}), (\ref{fin2})
and~(\ref{emident}) are independent of any physical constants.
The particular transformation chosen is
\[
  f = \eta F, \ \ g = \eta G, \ \ 
  r = \frac{\hbar x}{\mu c}, \ \
  E = \lambda \mu c^2, \ \ V = \mu c^2 U
\]
where $\eta^2 = \mu^3 c^4 / \epsilon^2 \hbar^2$. These redefined
variables have the following dimensions in terms of length ($L$):
\[
  [x] = L^0; \ \ [\lambda] = L^0;
  \ \ [F] = L^{-3/2}; \ \ [G] = L^{-3/2}; \ \ [U] = L^0.
\]
This yields the transformed equations:
\begin{eqnarray}
\label{ang1}
  \left[ \lambda - U(x) + 1 \right] F(x)
  - \left[ \frac{d G}{d x} + \frac{1+\kappa}{x} G(x) \right] &=& 0 \\
  \label{ang2}
  \left[ \lambda - U(x) - 1 \right] G(x)
  + \left[ \frac{d F}{d x} + \frac{1-\kappa}{x} F(x) \right] &=& 0 \\
  \label{ang3}
  \nabla^2 U + \left( F^2 + G^2 \right) &=& 0
\end{eqnarray}
where $\nabla^2$ is now the Laplacian with respect to the $x$ coordinate.
The mass of the soliton comes from the transformed version of the total
energy expression~(\ref{emident}),
\begin{equation}
\label{ang5}
  mc^2 = \frac{\hbar \mu c^3}{\epsilon^2}
  \left[ \lambda \int_0^\infty \left( F^2 + G^2 \right) \, x^2 \, d x 
  + \frac{1}{2} \int_0^\infty
  \left( \frac{d U}{d x} \right)^2 \, x^2 \,dx
  \right]
\end{equation}
and the total charge is given as the integral of the charge density
\begin{equation}
\label{ang4}
  e = \epsilon \int \rho \, d V_{ol}
  = \frac{\hbar c}{\epsilon} \int_0^\infty
  \left( F^2 + G^2 \right) \, x^2 \,d x.
\end{equation}

We will show that if the charge $\epsilon$ is replaced by $e$,
that the $f$ component of the spinor greatly dominates the $g$ component.
By substituting $\epsilon = e$ and choosing a value for $m$, the value
of $\mu$ can be determined numerically once the spatial extent of the
soliton is known.
In this case, the expectation value of the radius of the particle becomes
\[
  \langle r \rangle = \frac{\hbar c}{\mu c^2} \langle x \rangle
  = \frac{e^2}{\mu c^2} \frac{\hbar c}{e^2}
  \frac{\displaystyle \int \nabla^2 U x^3\,dx}{\displaystyle \int
  \nabla^2 U x^2\,dx}
  = \frac{e^2}{\mu c^2}
  \frac{\displaystyle \int \nabla^2 U x^3\,dx}{\displaystyle \left[ \int
  \nabla^2 U x^2\,dx\right]^2}.
\]
To stay within the current experimental bounds of the mean charge radius,
this value must be less than $r_{exp}$ which is $\le 10^{-18}m$ in the
case of an electron.  Hence,
\[
  \int_0^\infty \nabla^2 U x^3 \, dx \le \frac{r_{exp}}{r_e}
  \frac{\mu}{m_e} \left[ \int_0^\infty \nabla^2 U x^2\,dx \right]^2
\]
where $r_e$ is the classical electron radius $r_e = e^2/m_e c^2$.

Since we know that $U$ has zero slope at $x=0$ and that it must behave as
$N/x$ for large argument ($N$ is the amount of enclosed charge), we assume as a
first approximation, that $U$ can be represented as the electrostatic potential 
produced by a sphere of radius $R_0$ with uniform charge density.  Therefore,
\begin{equation}\label{uni}
  U(r) = \left\{ \begin{array}{l@{\quad \mbox{for} \quad} l}
         \displaystyle\frac{N}{R_0} \left[ \frac{3}{2}
         - \frac{1}{2} \frac{r^2}{R_0^2} \right] & r < R_0 \\
         \displaystyle\frac{N}{r} & r \ge R_0.
         \end{array} \right.
\end{equation}
With this representation, one finds that
\[
  \langle r \rangle = \frac{9}{4} \frac{r_e m_e R_0}{\mu N}
\]
which means that since $0 < \langle r \rangle < r_{exp}$, we can conclude that
\[
  0 < \frac{R_0}{\mu N} < \frac{4}{9} \frac{r_{exp}}{r_e m_e}
  \simeq 3.088 \times 10^{-4} \ c^2/\mbox{MeV}
\]
in the case of the electron.

Let $R_0$ be defined as the effective range of the non-Coulombic behaviour of
the potential energy so that for $x > R_0$, $U(x) \sim N/x$.  Since $U$ is
a solution to a Poisson equation with a negative definite charge density,
$U(0)$ must be larger than $U(R_0)$.  This can be quickly verified by
considering the opposite.  If $U(0) < U(R_0)$ then there exists some
$r \in (0,R_0)$ such that $U'(r) = 0$.  Therefore integrating~(\ref{ang3})
from $0$ to $r$, one obtains
\[
  r^2 U'(r) = 0 = - \int_0^r \left( F^2 + G^2 \right) x^2 \, dx
\]
which is clearly a contradiction.

To determine the initial values of $F$ and $G$, one simply eliminates either
$F$ or $G$ from~(\ref{ang1}-\ref{ang2}), say $F$, which leads to a second
order equation for the other, namely,
\[
  G'' + PG' + QG = 0
\]
where both $P$ and $Q$ are functions of $U$, $U'$, $\kappa$, $\lambda$ and $x$.
To avoid a singularity in the potential $U(x)$, it must be both bounded and
have zero slope in a neighbourhood of the origin. Moreover, both $F$ and $G$
are bounded in this same neighbourhood.  From this, it is easy to verify that
\begin{equation}\label{asdf1}
  F(0) = \left\{ \begin{array}{l@{\quad \quad} l}
            0 & \kappa = -1\\
            \mbox{arbitrary} & \kappa = +1\\
            0 & \forall \ \mbox{other} \ \kappa,
              \end{array} \right.
\end{equation}
\begin{equation}\label{asdf2}
  G(0) = \left\{ \begin{array}{l@{\quad \quad} l}
            \mbox{arbitrary} & \kappa = -1\\
            0 & \kappa = +1\\
            0 & \forall \ \mbox{other} \ \kappa.
              \end{array} \right.
\end{equation}
Furthermore, by examining the indicial equation, it can be shown that no
fractional powers exist in a power series solution of either $F$ or $G$
about the origin $x=0$.

Summarizing the boundary conditions:
\[
  U(x) < U(0) < \infty, \hspace{1cm}x \in [0,\infty),
\]
\[
  U'(0) = 0,
\]
together with the conditions~(\ref{asdf1}), (\ref{asdf2}).
For the case $\kappa = -1$, the initial values of $U$, $G$ and the 
energy $\lambda$ are determined by the requirement that the wave
function $\psi$, and hence both $F$ and $G$,
vanish exponentially as $x \to \infty$.


\section{Results}
\label{theresults}
In the search for numerical solutions it was specified that
$\kappa = -1$ and $\lambda = 1$ giving the set of differential
equations
\begin{eqnarray*}
  \frac{d G}{d x} &=& \left[ 2 - U(x) \right] F(x) \\
  \frac{d F}{d x} &=& - \frac{2}{x} F(x) + U(x) G(x) \\
  \nabla^2 U &=& - F^2 - G^2.
\end{eqnarray*}
To find a soliton, the values of $F(0)$, $G(0)$ are specified and
a search is made for the value of $U(0)$ whereby
$\lim_{x\to\infty} x F(x) = 0$ and $\lim_{x\to\infty} x G(x) = 0$.
Only values of $G(0)>0$ are considered because the equations are
symmetric under the transformation $G \to -G$, $F \to -F$, $U \to U$.
In a neighbourhood of a ground state soliton, the radial
probability density is numerically seen to have a single well-defined
minimum value for $x > 0$.
The choice of $\kappa = -1$ gives the initial condition $F(0) = 0$.

The choice of $\lambda = 1$ is simply a numerical convenience.
Outside the neighbourhood of a soliton it is expected that the potential
will behave as $U(x) \sim A + B/x$ for large $x$. The value of
$\lambda$ should have been chosen so that the asymptotic behaviour
of the potential $U(x)$ is purely Coulombic in nature. By defining
a shifted potential $\tilde{U}(x) = U(x) - \lim_{x\to\infty} U(x)$,
this value of $\lambda$ must satisfy $1-U(x) = \lambda - \tilde{U}(x)$.
Therefore after a soliton is found the value of $\lambda$ is given as
$\lambda = 1-\lim_{x\to\infty} U(x)$. In addition, the starting value of
$\tilde{U}(x)$ is given by $\tilde{U}(0) = U(0) + \lambda - 1$.

Using the redefined value of $\lambda$
the observed charge and mass of the particle are compared to the
values used in the Lagrangian by using
the expressions~(\ref{ang4}) and~(\ref{ang5}) respectively.
By defining
\[
  \begin{array}{c}
  {\cal P} = \displaystyle \int_0^\infty \left(F^2 + G^2\right) x^2 \, d x,
  \hspace{1cm}
  {\cal X} = \displaystyle\int_0^\infty \left(F^2 + G^2\right) x^3 \, d x, \\[2mm]
  {\cal E} = \lambda {\cal P} + \displaystyle \frac{1}{2}
  \displaystyle \int_0^\infty
  \displaystyle \left(\frac{d U}{d x}\right)^2 x^2 \, d x,
  \end{array}
\]
the charge ratio $\epsilon/e$ is given as
\[
  \frac{\epsilon}{e} = \frac{\hbar c}{e^2}{\cal P} = \frac{{\cal P}}{\alpha}
\]
where $\alpha$ is the fine structure constant. The mass ratio
$\mu/m = {\cal P}^2/\alpha {\cal E}$ and the expectation value for the
radius of the soliton is
\[
  \langle r \rangle = 
  \frac{\int (f^2+g^2)r^3 \, d r}{\int (f^2+g^2)r^2 \, d r} =
  \frac{\hbar}{\mu c}\frac{\int (F^2+G^2)x^3 \, d x}{\int (F^2+G^2)x^2 \, d x} =
  r_e \left(\frac{m_e}{m}\right) \frac{{\cal E}{\cal X}}{{\cal P}^3}.
\]

Both of the quantities ${\cal P}$ and ${\cal X}$ are positive. However,
depending upon the value of $\lambda$, ${\cal E}$ could be positive, negative
or even zero if the electromagnetic and ``bare mass'' terms in the energy
exactly cancel. A negative value for ${\cal E}$ will give an unphysical negative
value for the observed radius $\langle r \rangle$. Because of this ambiguity, both
the value of $\langle r \rangle$ and the particle width $\Delta r =
\sqrt{\langle r^2 \rangle - \langle r \rangle^2}$ are presented.
Tables~\ref{resultstbl} and~\ref{resultstbl2} respectively list the numerical
parameters and the observed properties of a number
of ground state particles found where $m$ was taken to be the observed mass
of the electron $m_e$.

\begin{figure}[b]
\setlength{\unitlength}{1pt}
\begin{picture}(0,0)(0,120)
\centerline{\psfig{figure=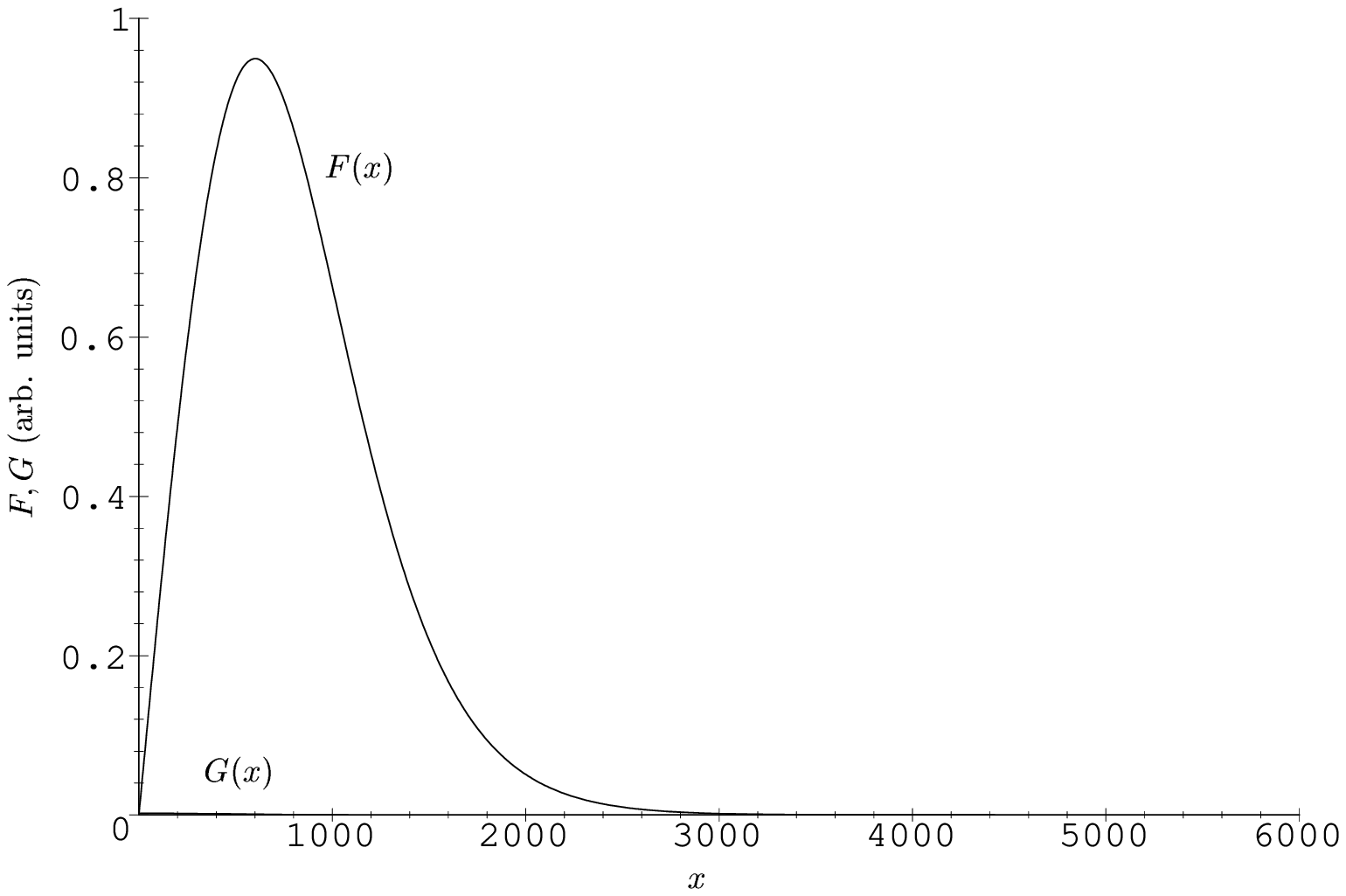}}
\end{picture}
\caption{\label{solfig1} $F$ and $G$ components of the wave function
for the case $\epsilon/e = 1$.} 
\vspace{4pt}
\small
  {\it Shown here is the radial dependence of the $F$ and $G$ components of 
  the soliton. Note that $G$ is much smaller than $F$. $G$ is barely
  discernible above the $x$ axis.}
\normalsize
\end{figure}

Figure~\ref{solfig1} illustrates the radial behaviour of $F$ and $G$
for the case $\epsilon = e \ (i = 1)$. It is to be noted that for $x > 0$,
$F$ is much larger than $G$ and as a consequence, $F G \ll F^2 + G^2$. In fact,
$G$ is so small that it resembles a straight line along the $x$ axis. This
supports the argument that the four-vector potential can be reasonably approximated
with only a radial $A^0$ component.

\begin{figure}[b]
\setlength{\unitlength}{1pt}
\begin{picture}(0,0)(0,120)
\centerline{\psfig{figure=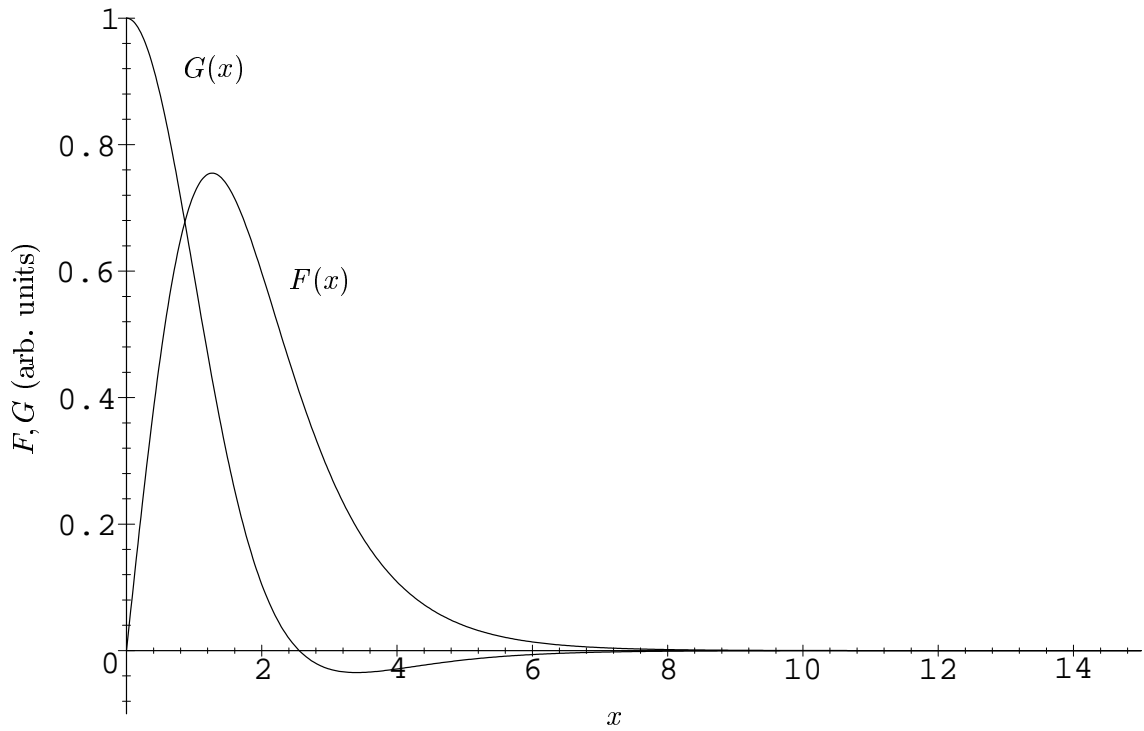}}
\end{picture}
\caption{\label{solfig2} $F$ and $G$ components of the wave function
for the case $\epsilon/e = 454.8$.} 
\vspace{4pt}
\small
  {\it Shown here is the radial dependence of the $F$ and $G$ components of 
  the soliton. The magnitudes of $F$ and $G$ are now comparable in contrast
  to the case when $\epsilon = e$.}
\normalsize
\end{figure}

The characteristics of a typical soliton with $\epsilon \ne e$ is
illustrated with the choice $\epsilon/e = 454.8 \ (i = 19)$.
In this case the potential plays a much more dominant role in holding the
particle together than in the case $\epsilon = e$. However, since in this case
the approximation of $FG \ll F^2 + G^2$ is violated, one would have to solve
the full model (equations~(\ref{sno6})-(\ref{sno7}))
to properly analyse this situation. This would be a far more complicated
problem. Figure~\ref{solfig2} illustrates the radial components of this spinor and
it shows that the magnitude of $G$ is now comparable to the magnitude
of $F$.  Table \ref{resultstbl2} also shows that the choice of
$\epsilon = 389.0 e$, $\mu = 2.360 \times 10^{12} m_e$ ($i=15$) yields
a soliton with an expectation value for the radius of $5.05 \times 10^{-23} m$.
This size is well within the present experimentally determined upper limit
for the electron radius of $\simeq 10^{-18} m$.

\begin{figure}[b]
\setlength{\unitlength}{1pt}
\begin{picture}(0,0)(0,120)
\centerline{\psfig{figure=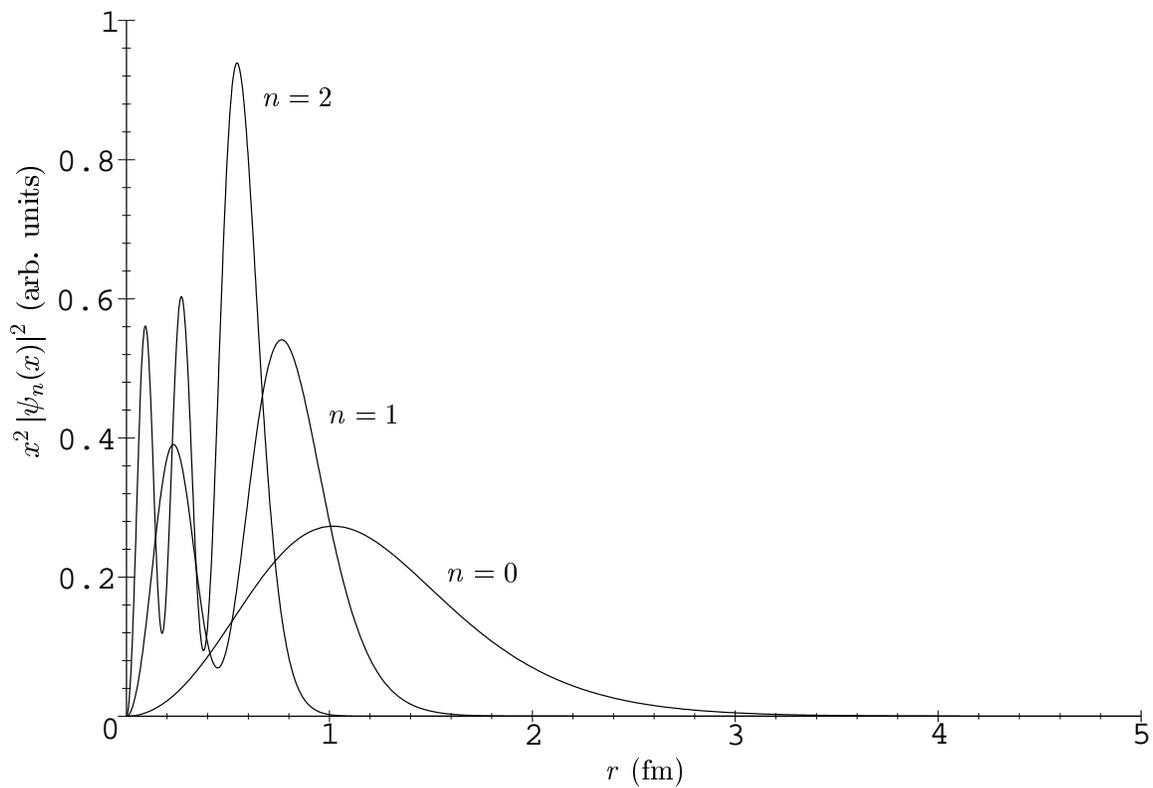}}
\end{picture}
\caption{\label{solfig3} Excited states of the theory.} 
\vspace{4pt}
\small
  {\it This figure shows the radial probability density of the first three
  particle state in the case $G(0)=1$. These particles have different charge
  ratios $\epsilon/e$. There exist excited states beyond the ones illustrated.}
\normalsize
\end{figure}

These equations also exhibit excited states.
The $n^{\mbox{\scriptsize th}}$ excited state of our field is characterized
through the functions $F_n(x)$, $G_n(x)$ and $U_n(x)$ for which the $G_n$
component crosses the abscissa $n+1$ times while the $F_n$ component crosses it
$n$ times. Once the ground state solution
is found, the value of $\mu$ can be determined through equation~(\ref{ang4}).
The corresponding $n^{\mbox{\scriptsize th}}$ excited state is that excited
state with the same observed charge ratio, $\epsilon/e$, as the ground state.
Therefore, in this interpretation of the theory, the ratio of the mass of the
$n^{\mbox{\scriptsize th}}$ excited state to the ground state is given by
the expression
\[
  \frac{m_n}{m_0} = \frac{\mu/m_0}{\mu/m_n} = \displaystyle
  \frac{\displaystyle\lambda_n \int_0^\infty \left(F_n^2+G_n^2\right) x^2 \, d x
  + \frac{1}{2} \int_0^\infty \left(\frac{d U_n}{d x}\right)^2 x^2 \, d x}
  {\displaystyle\lambda \int_0^\infty \left(F^2+G^2\right) x^2 \, d x
  + \frac{1}{2} \int_0^\infty \left(\frac{d U}{d x}\right)^2 x^2 \, d x}.
\]
Figure~\ref{solfig3} shows radial
probability density of the first three states for the case $G(0)=1$. Each of
these solitons has a different value of $\epsilon/e$.

\begin{figure}[b]
\setlength{\unitlength}{1pt}
\begin{picture}(0,0)(0,120)
\centerline{\psfig{figure=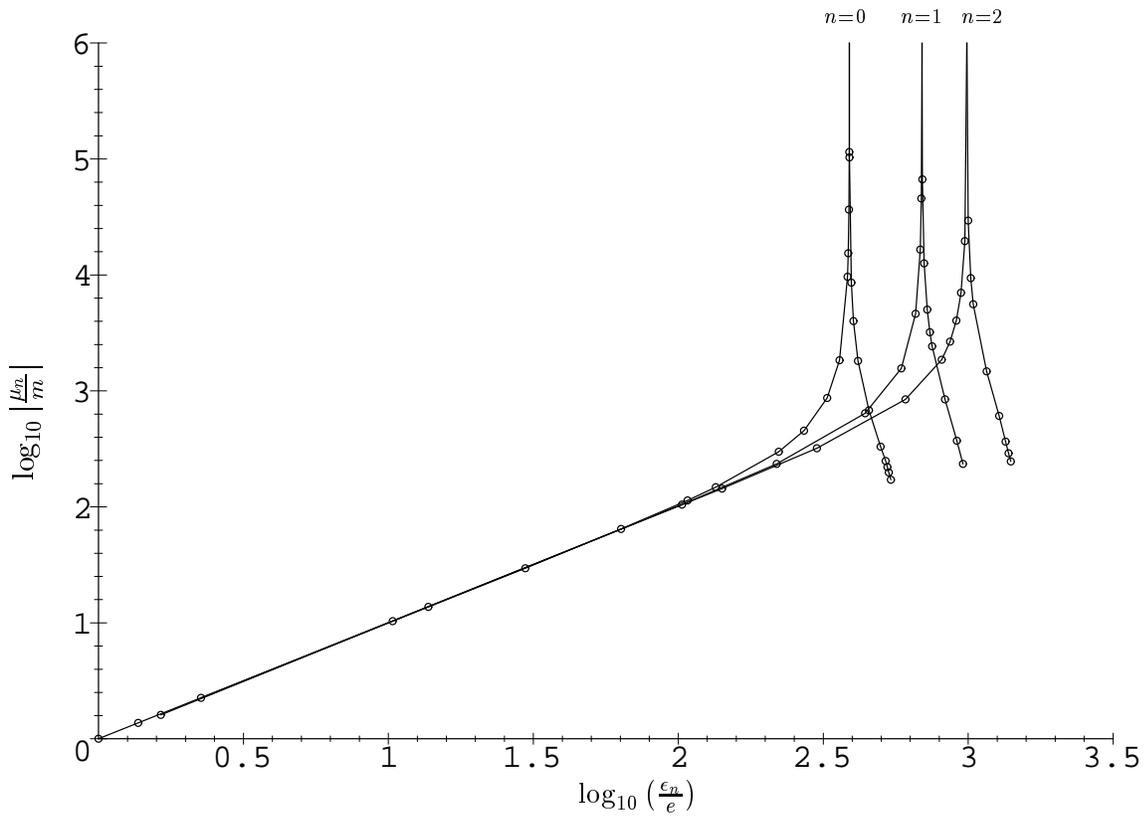}}
\end{picture}
\caption{\label{solfig4} Dependence of the mass ratio as a function
of the charge ratio.} 
\vspace{4pt}
\small
  {\it Shown is the dependence of the mass ratio} $\mu/m$ {\it as
  a function of the charge ratio} $\epsilon/e$ {\it for the ground
  state and first two excited state solitons. For each class of particles,
  there is a maximum charge ratio beyond which no solitons were found.}
\normalsize
\end{figure}

Figure~\ref{solfig4} illustrates the behaviour of the mass ratio, $\mu/m$, as
a function of the charge ratio $\epsilon/e$ for the ground state and the
first two excited states. For each class of particles there is a charge ratio
where the electromagnetic and bare mass components of the energy balance
making ${\cal E} = 0$. At this value of $\epsilon/e$, the mass ratio
$\mu/m \to \infty$. At charge ratios less than this critical value the
mass ratio is negative whereas charge ratios above this critical value
result in a positive value of $\mu/m$. There is numerical evidence that each
class of particles has an upper bound for the charge ratio. Above this maximum
charge ratio we were unable to find any solutions such that
$\lim_{x\to\infty} x F(x) \to 0$ or $\lim_{x\to\infty} x G(x) \to 0$. This
necessarily restricts the definition of the mass ratio defined above.
Figure~\ref{solfig4} also illustrates the fact that at moderate charge
ratios, the electromagnetic field does not contain an appreciable amount
of the particle energy resulting in the behaviour $|\mu/m| \simeq \epsilon/e$.

\begin{figure}[b]
\setlength{\unitlength}{1pt}
\begin{picture}(0,0)(0,120)
\centerline{\psfig{figure=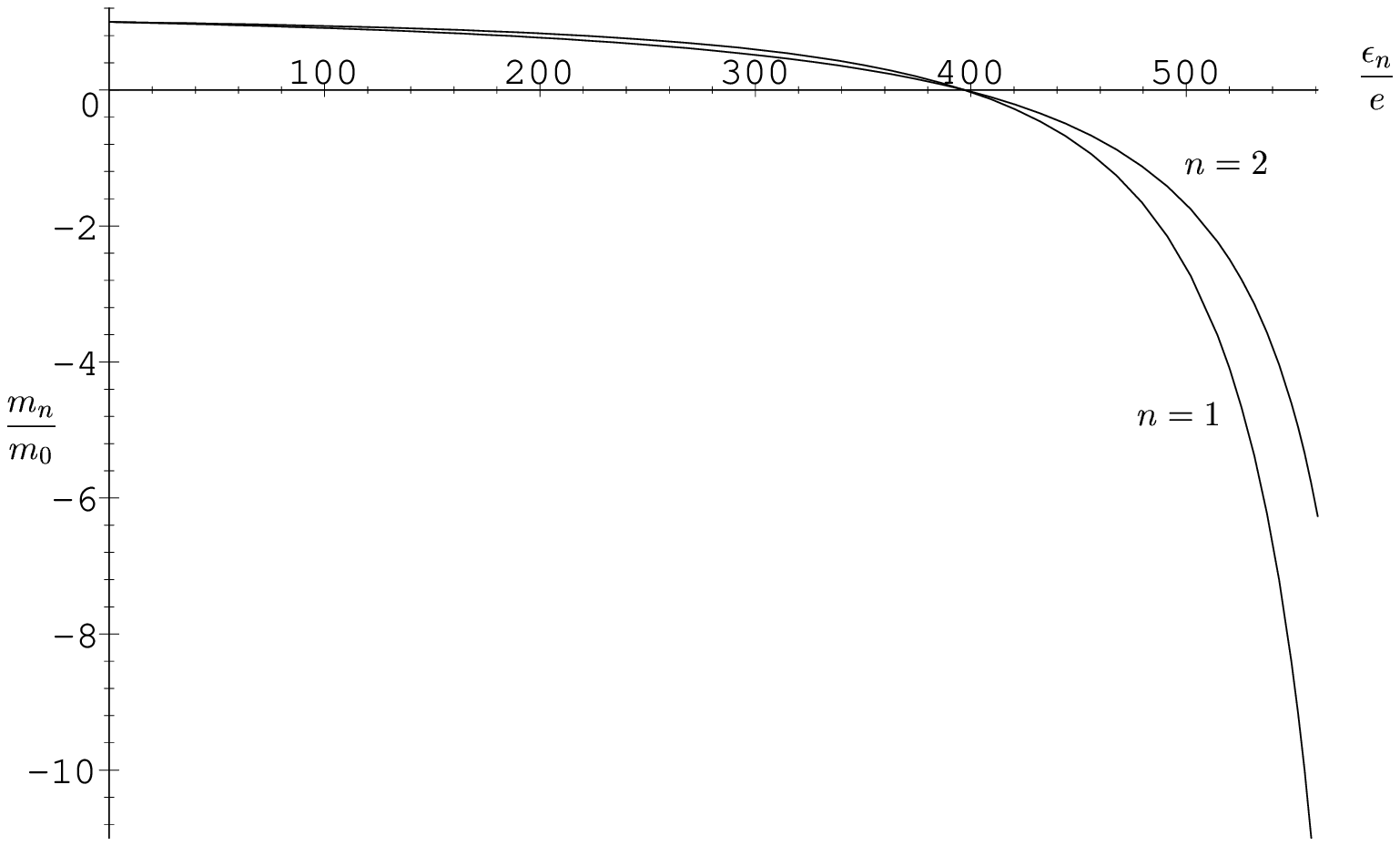}}
\end{picture}
\caption{\label{solfig5} Mass ratios of the first and second excited states
with respect to the ground state.} 
\vspace{4pt}
\small
  {\it There are essentially two regions of interest. For moderate charge
  ratios, the value of} $m_n/m_0 \simeq 1$ {\it with the mass ratio of the
  excited state} $n=2$ {\it slightly larger than for the} $n=1$ {\it state. Beyond
  the point where the ground state mass ratio becomes unbounded, the mass
  ratios begin to split. In this region, the} $|m_1/m_0|$ {\it ratio
  exceeds the} $|m_2/m_0|$ {\it ratio. In this region the approximation}
  $F G \ll F^2 + G^2$ {\it is no longer valid.}
\normalsize
\end{figure}

The mass ratios of the first and second excited states with respect to the
ground state solutions are shown in figure~\ref{solfig5}. This ratio is
only defined up to a maximum value of $\epsilon/e$ since beyond
$\epsilon/e \simeq 550$, a ground state fails to exist. 
For excited states, this maximum admissible
charge ratio increases. This implies that for a fixed value of $\epsilon/e$
there may not exist a ground state solution, but there will be
arbitrarily many excited states. As is readily apparent from
figure~\ref{solfig5}, the only appreciable mass splitting occurs for large
charge ratios. However, it is precisely for large charge ratios where
our approximation that $F G \ll F^2 + G^2$ breaks down.


\section{Concluding Remarks}
\label{concremarks}
We have seen that spherically symmetric Dirac-Maxwell solitons
can be constructed and with a charge and mass to model the electron
successfully. However, it should be noted that the higher energy excited
states of this form did not yield the large mass separations of the muon
and tau relative to the electron in this model. The search thus far has 
been restricted to spherical solitons. It is conceivable that a relaxation
of this restriction or some other change in conditions would increase 
the mass splitting. In any event, we have shown that Dirac-Maxwell solitons
exist and are capable of modelling an electron where the charge-to-mass
ratio is the observed $\simeq 10^{21}$ in units in which $G=c=1$.
Furthermore, we have found a charge-to-mass ratio that simultaneously
yields the observed charge and mass of the electron as well as exhibiting
a degreee of compactification that is well within the current experimental
upper limit.
Finster et al.\cite{finster} have
considered Einstein--Dirac--Maxwell (EDM) solitons and concluded that it is
the interaction with gravitation which is responsible for the existence of
bound states. However, we see here that bound states exist with negligible
gravitational interaction. While the $e/m$ ratio at which significant
gravitational coupling sets in is yet to be determined for EDM solitons,
it is our conjecture that this will be so at the same level that was found
earlier in the case of minimally coupled scalar
interaction\cite{cooperstock}, namely for $e/m \simeq 1$. The known
fundamental charged particles of nature, on the other hand
have enormous $e/m$ ratios.

\vspace{+40pt}

\appendix

\section{Derivatives of $f(r)Y^m_l(\theta,\varphi)$}
\setcounter{equation}{0}
\renewcommand{\theequation}{\Alph{section}.\arabic{equation}}
In the Dirac wave equation, all of the derivatives are with respect to
Cartesian coordinates.  We can change to a spherical polar representation
via the transformation
\[
  \begin{array}{l}
  x = r \sin \theta \, \cos \varphi \\
  y = r \sin \theta \, \sin \varphi \\
  z = r \cos \theta.\end{array}
\]
By applying the chain rule, it is trivial to show that this changes the
first order partial derivatives via
\begin{eqnarray}
  \frac{\partial}{\partial x}
  &=& \sin\theta \, \cos\varphi \, \frac{\partial}{\partial r} +
  \frac{\cos\theta\,\cos\varphi}{r}
  \frac{\partial}{\partial \theta} -
  \frac{\sin\varphi}{r\,\sin\theta} \frac{\partial}{\partial \varphi} \\[2mm]
  \frac{\partial}{\partial y}
  &=& \sin\theta \, \sin\varphi \, \frac{\partial}{\partial r} +
  \frac{\cos\theta\,\sin\varphi}{r}
  \frac{\partial}{\partial \theta} +
  \frac{\cos\varphi}{r\,\sin\theta} \frac{\partial}{\partial \varphi} \\[2mm]
  \frac{\partial}{\partial z}
  &=& \cos\theta\,\frac{\partial}{\partial r} - \frac{\sin\theta}{r}
  \frac{\partial}{\partial \theta}.
\end{eqnarray}
If the functions $\psi_j (j = 1,\ldots,4)$ from expression~(\ref{thestructure})
are substituted into~(\ref{starray})-(\ref{earray}), and if one uses the
formulas given in Bethe and Salpeter\cite{bands} for the derivatives of a
function of the form $f(r) Y_l^m(\theta,\varphi)$ with respect to $x$, $y$,
and $z$, one finds a coupled pair of first order ordinary equations for
$f(r)$ and $g(r)$.
	
For example, in order to calculate
\[
  \frac{\partial}{\partial z} \left[f(r) Y_l^m (\theta,\varphi)\right] ,
\]
we first require the identities
\begin{equation}
  \label{ident1}
  \cos\theta P_l^m(\cos\theta) = \frac{1}{2l+1} \left[(l-m+1)
  P_{l+1}^m(\cos\theta) 
  + (l+m) P_{l-1}^m(\cos\theta) \right]
\end{equation}
\begin{equation}
  \label{ident2}
  \sin\theta \frac{d}{d\theta} P_l^m(\cos\theta) =
  \frac{1}{2l+1} \left[l(l-m+1) P_{l+1}^m(\cos\theta)
  - (l+1)(l+m) P_{l-1}^m(\cos\theta)\right],
\end{equation}
which can both be verified through the use of Rodrigues' formula
\[
  P_l^m(\mu) = \frac{(-1)^m}{2^l l!} (1-\mu^2)^{m/2}
  \frac{d^{l+m}}{d\mu^{l+m}} (\mu^2-1)^l.
\]
Writing $Y_l^m(\theta,\varphi)$ as a function of $P_l^m$ by
using~(\ref{sphereharm}) gives the relationship
\[
  \frac{\partial}{\partial z} \left[ f(r) Y_l^m(\theta,\varphi) \right]
  = \sqrt{\displaystyle\frac{2l+1}{4\pi}\frac{(l-m)!}{(l+m)!}}
  \ e^{im\varphi} 
  \left[ \cos\theta\, P_l^m(\cos\theta) \frac{df}{dr} - \sin\theta\,
  \frac{d}{d\theta} P_l^m(\cos\theta) \frac{f}{r} \right].
\]
By substituting~(\ref{ident1}-\ref{ident2}) in the above, collecting terms,
and applying the definition of $Y_l^m(\theta,\varphi)$ once
again, one obtains the simplification
\begin{eqnarray}
  \frac{\partial}{\partial z} \left[ f(r) Y_l^m(\theta,\varphi) \right] &=&
  \sqrt{\displaystyle\frac{(l-m+1)(l+m+1)}
  {(2l+1)(2l+3)}} Y_{l+1}^m(\theta,\varphi)
  \left[\frac{df}{r} - \frac{l}{r} f \right] \nonumber \\
\label{bad1}
  &~& + \sqrt{\displaystyle\frac{(l-m)(l+m)}{(2l-1)(2l+1)}}
  Y_{l-1}^m(\theta,\varphi) \left[\frac{df}{d r} 
  + \frac{l+1}{r} f \right].
\end{eqnarray}
Similar relationships for
$\frac{\partial}{\partial x} \pm i \frac{\partial}{\partial y}$
can be found in Bethe and Salpeter\footnote{See formula (A.38) and (A.39)
respectively in Bethe and Salpeter.}, but there is a very 
elegant way to derive these operators by applying the Wigner--Eckart theorem.

First, we evaluate the matrix element
$\langle l \  0 | \nabla_0 | l \ 0 \rangle$ of the gradient operator,
which is an example of a vector operator.  Specifically,
\[
  \nabla_0 = \frac{\partial}{\partial z}, \ \ \
  \nabla_{\pm} = \mp \frac{1}{\sqrt{2}} \left( \frac{\partial}{\partial x} 
  \pm i \frac{\partial}{\partial y} \right).
\]
Since
\[
  \nabla_0 f(r) Y_l^0 = \frac{l+1}{\sqrt{(2l+1)(2l+3)}} Y_{l+1}^0
  \left[\frac{df}{dr} - \frac{l}{r} f \right] +
  \frac{l}{\sqrt{(2l-1)(2l+1)}} Y_{l-1}^0
  \left[\frac{df}{dr} + \frac{l+1}{r} f \right]
\]
\\
for the special case of~(\ref{bad1}) where $m = 0$, we have
\[
  \langle l' \  0 | \nabla_0 | l \ 0 \rangle =
  \frac{l+1}{\sqrt{(2l+1)(2l+3)}} \left[\frac{df}{dr} - \frac{l}{r} f \right] 
  \!\! \delta_{l+1}^{l'} + \frac{l}{\sqrt{(2l-1)(2l+1)}}
  \left[\frac{df}{dr} + \frac{l+1}{r} f \right] \!\! \delta_{l-1}^{l'}.
\]
Now, we are at a point where we can use the Wigner--Eckart theorem.
By inspection, the general matrix element is given by
\begin{eqnarray*}
  \langle l' \ m' | \nabla_{\mu} | l \ m \rangle &=& (-1)^{l'-m'}
  \left( \begin{array}{ccc}
  l' & 1 & l \\
  -m' & \mu & m
  \end{array}
  \right)
  \langle l' || \nabla || l \rangle \\
  &=& (-1)^{m'} \frac{
  \left( \begin{array}{ccc}
  l' & 1 & l \\
  -m' & \mu & m
  \end{array} \right)} {
  \left( \begin{array}{ccc}
  l' & 1 & l \\
  0 & 0 & 0
  \end{array} \right)}
  \langle l' \ 0 | \nabla_0 | l \ 0 \rangle.
\end{eqnarray*}
After evaluating the $3-j$ symbols, one can quickly verify the following
equations.
\begin{eqnarray}
  \frac{\partial}{\partial z} \left[(f(r)Y_l^m(\theta,\varphi) \right]
  &=& \sqrt{\frac{(l-m+1)(l+m+1)}{(2l+1)(2l+3)}}
  Y_{l+1}^m(\theta,\varphi) \left[ \frac{df}{dr}
  - \frac{l}{r} f \right] \nonumber \\
  \label{dbm0}
  &+& \!\!\!\!\! \sqrt{\frac{(l-m)(l+m)}{(2l-1)(2l+1)}}
  Y_{l-1}^m(\theta,\varphi) \left[ \frac{df}{dr} + \frac{l+1}{r} f \right] \\
  \left[\frac{\partial}{\partial x} + i \frac{\partial}{\partial y} \right]
  \left[f(r)Y_l^m(\theta,\varphi)\right]
  &=& \sqrt{\frac{(l+m+1)(l+m+2)}{(2l+1)(2l+3)}} 
  Y_{l+1}^{m+1}(\theta,\varphi) \left[ \frac{df}{dr}
  - \frac{l}{r} f  \right] \nonumber \\
  \label{dbm1}
  &-& \!\!\!\!\! \sqrt{\frac{(l-m-1)(l-m)}{(2l-1)(2l+1)}}
  Y_{l-1}^{m+1}(\theta,\varphi) \left[ \frac{df}{dr} + \frac{l+1}{r} f \right] \\ 
  \left[\frac{\partial}{\partial x} - i \frac{\partial}{\partial y} \right] 
  \left[ f(r)Y_l^m(\theta,\varphi) \right]
  &=& -\sqrt{\frac{(l-m+1)(l-m+2)}{(2l+1)(2l+3)}} 
  Y_{l+1}^{m-1}(\theta,\varphi) \left[ \frac{df}{dr}
  - \frac{l}{r} f \right] \nonumber \\
  \label{dbm2}
  &+& \!\!\!\!\! \sqrt{\frac{(l+m-1)(l+m)}{(2l-1)(2l+1)}} Y_{l-1}^{m-1}(\theta,\varphi) 
  \left[ \frac{df}{dr} + \frac{l+1}{r} f \right].
\end{eqnarray}
Linear combinations of~(\ref{dbm1}) and~(\ref{dbm2}) yield the derivatives
with respect to $x$ and $y$.

\newpage

\begin{table}[t]     
\unitlength1.0cm
\begin{center}
\begin{tabular}{|cc|}
  \hline
   States $|l',m',j'\rangle$ & Corresponding Potential $V(\boldr)$ \\
  \hline
  \hline
   \ & \ \\ [-0.4cm]
   $|0,0,1/2\rangle, |0,-1,1/2\rangle$ & $e^2 I_0$ \\ [0.2cm]
   $|1,0,1/2\rangle, |1,-1,1/2\rangle$ & $e^2 I_0$ \\ [0.2cm]
   $|1,1,3/2\rangle, |0,-2,3/2\rangle$ & $e^2
   \left[ I_0 - \frac{3}{2} \left( 3\cos^2\theta' - 1\right) I_2\right]$ \\ [0.2cm]
   $|1,0,3/2\rangle, |0,-1,3/2\rangle$ & $e^2
   \left[ I_0 + \frac{3}{2} \left( 3\cos^2\theta' - 1\right) I_2\right]$ \\ [0.2cm]
  \hline
\end{tabular}
\end{center}
\caption{\label{tab3} The Dirac--Maxwell particle self-field potential.}
\small
  {\it The self-field potential energy for a Dirac--Maxwell particle in the
  states $l=0,1$ where
  \[
    I_l(r) = \displaystyle \int_0^r \left[ f(r')^2 + g(r')^2 \right]\,
    \frac{r'^{l+2}}{r^{l+1}}\,d r'
    + \int_r^{\infty} \left[ f(r')^2 + g(r')^2 \right]\,
    \frac{r^l}{r'^{l-2}}\,d r'.
  \]}
\normalsize
\drline
\end{table}
~

\newpage
\begin{table}[!t] 
\unitlength1.0cm
\begin{center}
\begin{tabular}{|c|c|c|c|c|}
  \hline
  $i$ & $G(0)$ & $U(0)$ & $\lambda$ & $x_{\max}$ \\
  \hline
  \hline
1  &$3.864 \times 10^{-9}$ & $4.8879852 \times 10^{-6}$ & $-0.99999712$ & 8342.9 \\ %
2  &  $1.0 \times 10^{-8}$ & $9.2122047 \times 10^{-6}$ & $-0.99999457$ & 6103.6 \\ %
3  &  $1.0 \times 10^{-5}$ & $9.2122842 \times 10^{-4}$ & $-0.99945701$ & 688.37 \\ %
4  &  $1.0 \times 10^{-4}$ & $4.2760987 \times 10^{-3}$ & $-0.99748078$ & 330.61 \\ %
5  &  $1.0 \times 10^{-3}$ & $1.9850763 \times 10^{-2}$ & $-0.98833078$ & 158.92 \\ %
6  &  $5.0 \times 10^{-3}$ & $5.8063758 \times 10^{-2}$ & $-0.96604870$ & 81.570 \\ %
7  &  $1.0 \times 10^{-2}$ & $9.2196988 \times 10^{-2}$ & $-0.94634242$ & 77.061 \\
8  &  $5.0 \times 10^{-2}$ & $2.6992683 \times 10^{-1}$ & $-0.84654287$ & 50.139 \\
9  &  $1.0 \times 10^{-1}$ & $4.2884212 \times 10^{-1}$ & $-0.76093227$ & 38.686 \\
10 &  $2.0 \times 10^{-1}$ & $6.8134130 \times 10^{-1}$ & $-0.63095887$ & 33.509 \\
11 &  $3.0 \times 10^{-1}$ & $8.9315883 \times 10^{-1}$ & $-0.52683528$ & 23.495 \\
12 &  $4.0 \times 10^{-1}$ & $1.0821393 \times 10^{0}$  & $-0.43716613$ & 26.964 \\
13 &  $4.1 \times 10^{-1}$ & $1.1001061 \times 10^{0}$  & $-0.42878460$ & 27.544 \\
14 &  $4.2 \times 10^{-1}$ & $1.1179261 \times 10^{0}$  & $-0.42049493$ & 26.049 \\

15 &  $G^*(0)$             & $1.1309487 \times 10^{0}$  & $-0.41445154$ & 24.146 \\
16 &  $4.3 \times 10^{-1}$ & $1.1356039 \times 10^{0}$  & $-0.41229414$ & 24.382 \\
17 &  $5.0 \times 10^{-1}$ & $1.2557044 \times 10^{0}$  & $-0.35715757$ & 24.156 \\
18 &  $6.0 \times 10^{-1}$ & $1.4178526 \times 10^{0}$  & $-0.28421512$ & 22.584 \\
19 &  $1.0 \times 10^{0}$  & $1.9913670 \times 10^{0}$  & $-0.03776277$ & 20.438 \\
20 &  $2.0 \times 10^{0}$  & $3.1519761 \times 10^{0}$  & $+0.42244841$ & 18.125 \\
  \hline
\end{tabular}
\end{center}
\caption{\label{resultstbl} Numerical parameters for a set of various
ground state particles.}
\small
  {\it For each particle, the value of} $G(0)$ {\it is selected and one
  searches for the value of} $U(0) + \lambda$ {\it that gives a bounded
  solution. The physical parameters are computed from the solution
  defined on} $x \in [0,x_{\max}]$.  $G^*(0) = 0.4273589430$.
\normalsize
\drline
\end{table}
~

\newpage

\begin{table}[!t] 
\unitlength1.0cm
\begin{center}
\begin{tabular}{|c|c|c|c|c|c|}
  \hline
  $i$ & $\epsilon/e$ & $\mu/m$ & $\langle r \rangle$ & $\Delta r$ \\
  \hline
  \hline
1  &  1.000 & $-1.000 \times 10^{0}$ & $-6.61 \times 10^{-8}$  & $2.31 \times 10^{-8}$  \\ %
2  &  1.371 & $-1.371 \times 10^{0}$ & $-3.51 \times 10^{-8}$  & $1.22 \times 10^{-8}$  \\ %
3  &  13.74 & $-1.374 \times 10^{1}$ & $-3.51 \times 10^{-10}$ & $1.22 \times 10^{-10}$ \\ %
4  &  29.55 & $-2.955 \times 10^{1}$ & $-7.53 \times 10^{-11}$ & $2.63 \times 10^{-11}$ \\ %
5  &  63.46 & $-6.462 \times 10^{1}$ & $-1.60 \times 10^{-11}$ & $5.61 \times 10^{-12}$ \\ %
6  &  107.6 & $-1.135 \times 10^{2}$ & $-5.30 \times 10^{-12}$ & $1.87 \times 10^{-12}$ \\ %
7  &  134.7 & $-1.479 \times 10^{2}$ & $-3.21 \times 10^{-12}$ & $1.14 \times 10^{-12}$ \\
8  &  222.1 & $-2.991 \times 10^{2}$ & $-9.07 \times 10^{-13}$ & $3.31 \times 10^{-13}$ \\
9  &  271.3 & $-4.504 \times 10^{2}$ & $-4.64 \times 10^{-13}$ & $1.73 \times 10^{-13}$ \\
10 &  326.2 & $-8.687 \times 10^{2}$ & $-1.87 \times 10^{-13}$ & $7.23 \times 10^{-14}$ \\
11 &  359.7 & $-1.838 \times 10^{3}$ & $-7.51 \times 10^{-14}$ & $2.99 \times 10^{-14}$ \\
12 &  383.6 & $-9.658 \times 10^{3}$ & $-1.27 \times 10^{-14}$ & $5.16 \times 10^{-15}$ \\
13 &  385.6 & $-1.538 \times 10^{4}$ & $-7.89 \times 10^{-15}$ & $3.21 \times 10^{-15}$ \\
14 &  387.6 & $-3.666 \times 10^{4}$ & $-3.28 \times 10^{-15}$ & $1.34 \times 10^{-15}$ \\

15 &  389.0 & $+2.360 \times 10^{12}$& $+5.05 \times 10^{-23}$ & $2.06 \times 10^{-23}$ \\
16 &  389.5 & $+1.032 \times 10^{5}$ & $+1.15 \times 10^{-15}$ & $4.71 \times 10^{-16}$ \\
17 &  401.8 & $+3.998 \times 10^{3}$ & $+2.79 \times 10^{-14}$ & $1.16 \times 10^{-14}$ \\
18 &  416.4 & $+1.818 \times 10^{3}$ & $+5.68 \times 10^{-14}$ & $2.39 \times 10^{-14}$ \\
19 &  454.8 & $+6.800 \times 10^{2}$ & $+1.21 \times 10^{-13}$ & $5.35 \times 10^{-14}$ \\
20 &  498.7 & $+3.305 \times 10^{2}$ & $+1.78 \times 10^{-13}$ & $8.59 \times 10^{-14}$ \\
  \hline
\end{tabular}
\end{center}
\caption{\label{resultstbl2} Corresponding observable quantities for a set of various
ground state particles.}
\small
  {\it The values} $\langle r \rangle$ {\it and} $\Delta r$ {\it are
  measured in meters and are computed from a soliton defined on}
  $x \in [0,x_{\max}]$. {\it For this calculation it is assumed that}
  $m = m_e$.
\normalsize
\drline
\end{table}
~

\newpage

~

\newpage

~

\newpage

~

\newpage

~

\newpage

~


\begin{center}
  {\bf REFERENCES}
\end{center}
\begin{thebibliography}{9}

\vspace{-40pt}

\bibitem{einstein}
  Einstein, A. \& Rosen, N. (1935).
  Physical Review, {\bf 48}, 73-77.

\bibitem{rosen}
  Rosen, N. (1939).
  Physical Review, {\bf 55}, 94-101.

\bibitem{rosenstock}
  Rosen, N. \& Rosenstock, H. B. (1952).
  Physical Review, 
  {\bf 85}(2), 257-259.

\bibitem{cooperstock}
  Cooperstock, F. I. \& Rosen, N. (1989).
  International Journal of Theoretical 
  Physics, {\bf 28}(4), 423-440.
 
\bibitem{cooperstock2}
  Cooperstock, F. I. (1991). {\it The Electron: New Theory and Experiment},
  Eds. D. Hestenes and A. Weingartshofer, Kluwer Academic.
  
\bibitem{bohun}
  Bohun, C. S. (1991). {\it A Self-Consistent Dirac--Maxwell Field of
  Solitons}, MSc. Thesis, University of Victoria.

\bibitem{lisi}
  Lisi, A. G. (1995).
  Journal of Physics A: Mathematical and General, {\bf 28}, 5385-5392.
 
\bibitem{finster}
   Finster, F., Smoller, J. and Yau, S-T., preprint gr-qc/9801079

\bibitem{grif}
  Griffiths, D. J. (1987). {\it Introduction to Elementary Particles}.
  New York: Harper and Row.

\bibitem{bands}
  Bethe, H. A. \& Salpeter, E. E. (1957).
  {\it Quantum Mechanics of One- and Two-
  Electron Atoms}. Berlin: Springer-Verlag.

\bibitem{evans}
  Evans, L. C. (1998). {\it Partial Differential Equations: Graduate Studies in
  Mathematics, vol 19}.
  American Mathematical Society,
  Providence, Rhode Island. p. 22.

\bibitem{brpm}
  Brezzi, F. \& Markowich, P. A. (1991).
  Mathematical Methods in the Applied Sciences, {\bf 14}, 35-61. 

\bibitem{abram}
  Abramowitz, M \& Stegun, I. A. (1964). {\it
  Handbook of mathematical functions, with formulas, graphs, and
  mathematical tables}. National Bureau of Standards,
  United States Department of Commerce.

\bibitem{landl}
  Landau, L. D. \& Lifshitz E. M. (1971). {\it
  The classical theory of fields} (4th ed.). 
  New York: Pergamon Press.

\end{thebibliography}
\end{document}